\documentclass[prd,aps,preprint,nofootinbib]{revtex4}
\usepackage{epsfig}
\usepackage{amsmath}

\begin{document}

\preprint{RBRC-717} \preprint{LBNL-63751}

\title{Heavy Quarkonium Production in Single Transverse Polarized
High Energy Scattering}

\author{Feng Yuan}
\email{fyuan@lbl.gov} \affiliation{Nuclear Science Division,
Lawrence Berkeley National Laboratory, Berkeley, CA
94720}\affiliation{RIKEN BNL Research Center, Building 510A,
Brookhaven National Laboratory, Upton, NY 11973}

\begin{abstract}
We formulate the single transverse spin asymmetry in heavy quarkonium
production in lepton-nucleon and nucleon-nucleon collisions
in the non-relativistic limit. We find
that the asymmetry is very sensitive to the production mechanism. The final
state interactions with the heavy quark and antiquark cancel out among themselves when
the pair are produced in a color-single configuration, or cancel out with the initial
state interaction in $pp$ scattering when they are in color-octet. As a consequence, the asymmetry
is nonzero in $ep$ collisions only in the color-octet model, whereas in $pp$
collisions only in the color-singlet model.
\end{abstract}

\maketitle

\newcommand{\be}{\begin{equation}}
\newcommand{\ee}{\end{equation}}
\newcommand{\ben}{\[}
\newcommand{\een}{\]}
\newcommand{\beqn}{\begin{eqnarray}}
\newcommand{\eeqn}{\end{eqnarray}}
\newcommand{\Tr}{{\rm Tr} }

{\bf 1. Introduction.} Single-transverse spin asymmetry (SSA) is
a novel phenomena in hadronic reactions \cite{Anselmino:1994gn,{Qiu:1991pp}}, and has
long been observed in various processes.  In
these processes, a transversely polarized nucleon scatters off a
unpolarized nucleon (or virtual photon) target, and the final
observed hadron have asymmetric distribution in the
transverse plane perpendicular to the beam direction depending on
the polarization vector of the scattering nucleon. Recent
experimental studies have
motivated much theoretical developments for understanding the
underlying physics associated with the SSA phenomena.

An important theoretical progress made in the last few years
is the uncover of the crucial
role of the initial/final state interactions \cite{Brodsky:2002cx},
which leads to a non-vanishing SSA in the Bjorken limit in the
semi-inclusive hadron production in deep inelastic scattering (SIDIS) and Drell-Yan
lepton pair production processes. These initial/final state interactions were understood
as a result of the gauge link in the gauge invariant transverse
momentum dependent (TMD) quark distributions
\cite{Collins:2002kn,{Ji:2002aa},Boer:2003cm}, and
the SSA can be traced back to a
naive time-reversal-odd distribution, the so-called
Sivers function \cite{Sivers:1990fh}. The difference between the initial
and final state interaction in the above two processes results
into a sign change for the associated quark Sivers function and
the SSAs. The Sivers function has been extended to the gluon sector, and
the associated prospects have been investigated in recent years
\cite{Boer:2003tx,Burkardt:2004ur,Anselmino:2004nk,
Anselmino:2006yq,{Brodsky:2006ha},Bomhof:2006ra}.

Heavy quark and quarkonium productions are natural probes for
the gluon Sivers function~\cite{Anselmino:2004nk,liu}. Especially at low transverse momentum, their
production will be sensitive to the intrinsic transverse momentum~\cite{Berger:2004cc}, and
can be used to study the gluon Sivers effect.
Meanwhile, the heavy quarkonium production has attracted
many experimental and theoretical investigations in the last decade, starting from the
anomalous production discovered at the Tevatron $p\bar p$ experiment~\cite{Abe:1992ww}
and a theoretical framework for studying the heavy quarkonium system, the so-called non-relativistic
QCD (NRQCD)~\cite{Bodwin:1994jh}. The basic argument for the NRQCD factorization
is that the heavy quark pair are produced at short distance in a color-singlet or
color-octet configurations. The hadronization of the pair (in singlet or octet) is described by the
associated matrix elements, which can be
characterized according to the velocity expansion~\cite{Bodwin:1994jh}. This framework had
success in explaining some experimental observations.
Yet, we do not have a definitive answer for the production
mechanism~\cite{Brambilla:2004wf,Nayak:2005rw,Zhang:2005cha,{Campbell:2007ws}}.

In this paper, we formulate the SSA in heavy quarkonium production by carefully examining the
initial and final state interaction effects. We follow the general arguments of the NRQCD factorization. The
SSA depends on the
color configuration of the pair produced at short distance, although the hadronization details
are not relevant to obtaining a non-zero SSA. Therefore, the experimental study of the SSAs
shall shed light on the production mechanism for the heavy quarkonium.
We will focus on the gluon-gluon (photon-gluon) fusion contributions in
two processes: one is the lepton-nucleon $ep$ scattering and one is nucleon-nucleon $pp$ scattering,
schematically as
\begin{equation}
A(P_A,S_\perp)+B(P_B)\to [Q\bar Q]^{(1,8)}+X\to H+X \ ,
\end{equation}
where $A$ is the polarized nucleon with  momentum $P_A$, $B$ the un-polarized hadron
or photon with momentum $P_B$. The heavy quark pair $Q\bar Q$ produced at short
distance ($\sim 1/M_Q$, $M_Q$ the heavy quark mass) can be in a color-singlet (labeled with $(1)$)
or color-octet (labeled with $(8)$) configurations, and $H$ represents the final physical
quarkonium state. The SSAs coming from the initial/final state interactions are calculated
based on the interactions between the quark pair or the incident gluon with the remanet of the polarized
target. We are interested in a kinematic region of low transverse momentum for the heavy quarkonium,
$P_\perp\ll M_Q$, where the TMD parton distribution is relevant.
In the following, we will present a general analysis for the SSAs coming from these interactions
in this kinematic region.
A detailed argument for low $P_\perp$ factorization in terms
of the TMD parton distributions is also important. 
Especially, when $P_\perp$ is the same order of the heavy quark kinetic
energy in the quarkonium's rest frame, the hadronization of the heavy quark
pair might interfere with the initial gluon radiation and break the TMD
as well as NRQCD factorization ~\cite{Berger:2004cc,jwqiu}. We will come to this important issue in the future.

{\bf 2. Final state interactions with the quark pair in the non-relativistic limit.}
We first discuss how to formulate the final state interactions effects
with the quark pair, taking the lepton-nucleon scattering as an example.
In this process, the dominant subprocess is the photon-gluon fusion channel, where the
photon comes from the lepton radiation. As we discussed in the introduction, in order
to obtain a nonzero Sivers-type SSA, we have to generate a phase from either the initial or final
state interactions. Because the lepton (photon) is colorless, there
is no initial state interaction in this process.

\begin{figure}[t]
\begin{center}
\includegraphics[width=10cm]{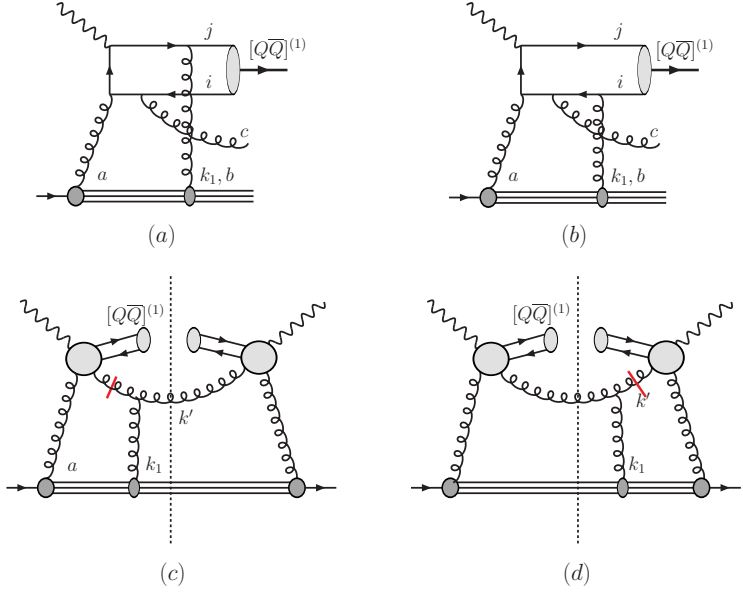}
\end{center}
\vskip -0.4cm \caption{\it Vanishing SSA in
photo(lepto)-production of heavy quarkonium when the heavy quark
pair are produced in a color-singlet configuration. The two final
state interactions with the quark (a) and antiquark (b) cancel
out each other; the final state interactions with the unobserved
particles cancel out among different cuts.} \label{fig1}
\end{figure}

In Figs.~1 and 2 we plot the generic final state interaction diagrams, for color-singlet and color-octet
cases, respectively. If the pair are produced in a color-single configuration, a gluon has to be
radiated. In general, we will have final state interactions with the quark pair (Fig.~1(a,b)) and
the radiated gluon (Fig.~1(c,d)). The interaction with the unobserved particle (here is the
radiated gluon) vanishes
after we summing different cut diagrams~\cite{Qiu:1991pp,{Koike:2007dg}}.
For example, the contribution from Fig.~1(c) is
proportional to pole contribution from the propagator labeled by a short bar with momentum
$k'+k_1$,
$$\frac{1}{(k'+k_1)^2+i\epsilon}\delta ((k')^2)|_{\rm pole}=-i\pi\delta ((k'+k_1)^2)\delta((k')^2)\ , $$
whereas the contribution from Fig.~1(d) will be
$$\frac{1}{(k')^2-i\epsilon}\delta ((k'+k_1)^2)|_{\rm pole}=+i\pi\delta ((k')^2)\delta((k'+k_1)^2)\ .$$
Clearly, these two contributions cancel out each other, because the other parts of the
scattering amplitudes for these two diagrams are the same except the above
pole contributions which are opposite to each other. The above result is quite general, and
shows that all the final state interactions with the unobserved particles do not contribute
to the SSA for the associated process.

Therefore, we only need to consider the final state interactions with the quark pair. Two example
diagrams are shown in Figs.~1(a) and (b). In the non-relativistic limit, the final state interaction with
the quark in Fig.~1(a) can be derived as follows,
\begin{eqnarray}
&&\Phi(k-\frac{P}{2};ij)(-ig)\gamma^\rho T^b i\frac{\not\!\! k-\not\!\!k_1+M_Q}{(k-k_1)^2-M_Q^2+i\epsilon}
\Gamma\nonumber\\
&&~~~~~~~\approx
\frac{g}{-k_1^++i\epsilon}\Phi(k-\frac{P}{2};ij)T^b\Gamma \ ,
\end{eqnarray}
where $k$ and $P$ are momenta for the quark and the quark pair, respectively, $\Phi$ represents
the wave function for the pair, $ij$ are color indices for the quark and antiquark, $b$ is the
color index for the gluon attaching to the quark line, $\rho$ is the index contracted with the
gauge potential $A^\rho$, and $\Gamma$ represents other
hard part for the scattering amplitude. The light-cone momentum components
are defined as $k^\pm=(k^0\pm k^z)/\sqrt{2}$,
and we assume that the polarized nucleon is moving along the $+\hat z$ direction:
$P_A=(P_A^+,0^-,0_\perp)$. In the non-relativistic limit, each of the quark and antiquark carries half of
the pair's momentum, $k\approx P/2$, and they are on mass shell: $k^2\approx M_Q^2$. More over,
the dominant contribution from the gluon interaction with the nucleon remanet comes from
the gauge potential $A^+$ in the covariant
gauge, and where the gluon momentum is collinear to
the polarized nucleon. In above derivation, we have also only taken the leading power contributions,
and neglected all higher order correction in terms of $1/M_Q$. This derivation shows that we can
simplify the final state interaction as an eikonal propagator.
Similarly, when the gluon attaches to the antiquark, Fig.~1(b), the contribution will be proportional to
\begin{equation}
\frac{-g}{-k_1^++i\epsilon}T^b\Phi(k-\frac{P}{2};ij)\Gamma \ ,
\end{equation}
where the $\Gamma$ is the same as above. The minus sign of this result comes
from the interaction with the antiquark. We can combine the contributions from
these two diagrams together, and it is proportional to
\begin{equation}
\frac{g}{-k_1^++i\epsilon}\left(\Phi(k-\frac{P}{2};ij)T^b-T^b\Phi(k-\frac{P}{2};ij)\right)\Gamma \ .
\end{equation}
From this, we immediate find that the SSA contribution vanishes if the pair are produced
in a color-singlet configuration, for which the color matrix for $\Phi$ will be $\Phi_{ij}^{(1)}\propto \delta_{ij}$,
and the above two terms cancel out each other completely. This observation is generic, and only
depends on the non-relativistic approximation we made and it is
valid for higher orders too. Therefore, we conclude that the SSA for heavy quarkonium
production in $ep$ scattering vanishes in the non-relativistic limit in the color-singlet
model.

\begin{figure}[t]
\begin{center}
\includegraphics[width=10cm]{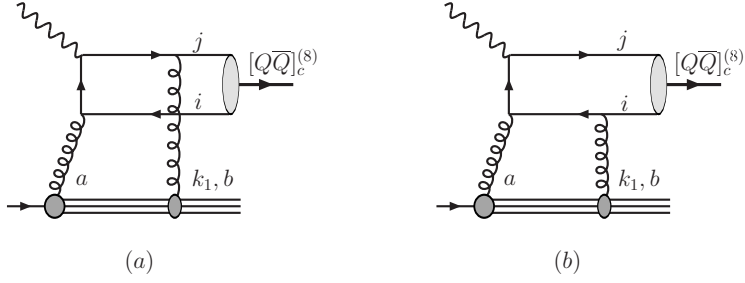}
\end{center}
\vskip -0.4cm \caption{\it Nonzero SSA in $ep$ collisions arise
when the heavy quark pair are produced in a color-octet
configuration. The two final state interactions with the quark
(a) and antiquark (b) have net effects.} \label{fig2}
\end{figure}
However, the above arguments do not hold when the pair are produced in a color-octet
configuration, for which we will have net effects from the two final state interactions.
We show these interactions in Fig.~2. Because the
pair are in color-octet, the color matrix of the wave function can be parameterized as
$\Phi^{(8,c)}_{ij}\propto T^c_{ij}$ where $c=1,\cdots,8$ represents the color-index of the pair.
The sum of the two final state interactions will be proportional to
\begin{equation}
\frac{g}{-k_1^++i\epsilon}(-if_{bcd})\Phi^{(8,d)}\Gamma \ ,
\end{equation}
where the latter factor is the same hard part in the above scattering amplitude. This result can
be summarized into a diagram shown in Fig.~3(a), and can be easily extended to
a two-gluon exchange contributions (b), which is proportional to
\begin{equation}
\frac{g}{-k_2^++i\epsilon}\frac{g}{-k_1^+-k_2^++i\epsilon}(-if_{dce})(-if_{bef})\Phi^{(8,f)}\Gamma\ .
\end{equation}
When it is generalized to all orders, we will find the final state interactions can
be summed into a gauge link associated with the gluon distribution, and the SSA depends
on the gluon Sivers function, which is the spin-dependent part of the following
distribution~\cite{Ji:2005nu},
\begin{eqnarray}
&&xG_{1T}^\perp(x,k_\perp)=\int\frac{d\xi^-d^2\xi_\perp}{P^+(2\pi)^3}
    e^{-ixP^+\xi^-+i\vec{k}_\perp\cdot \vec\xi_\perp}\label{e7}\\
    &&~~\times
\left\langle PS_\perp|{F^+}_\mu(\xi^-,\xi_\perp)
    {\cal L}^\dagger_{\xi^-,\xi_\perp} {\cal L}_{0,0_\perp}
F^{\mu+}(0)|P S_\perp\right\rangle \ , \nonumber
\end{eqnarray}
where sum over color indices is implicit.  $F^{\mu\nu}$ is the
gluon field strength tensor, $F_a^{\mu\nu}=\partial ^\mu A_a^\nu
-\partial^\nu A_a^\mu-g f_{abc} A^\mu_b A^\nu_c$,
$x$ the momentum fraction carried by the gluon,
$k_\perp$ the transverse momentum. ${\cal L}_\xi$  the
process-dependent gauge link~\cite{Collins:2002kn,{Ji:2002aa}}. For the diagrams in Fig.~3, the gauge link
sums all the final state interactions with the quark pair
(in color-octet state), for which we have future pointing gauge link
going to $+\infty$,
\begin{eqnarray}
  {\cal L}_{\xi^-,{\xi_\perp}}\to {\cal L}_\xi^\infty =
  P\exp\left(-ig\int^{\infty}_{0}
    d\zeta^- A^+(\zeta^- + \xi) \right) \ ,\label{fl}
\end{eqnarray}
in the covariant gauge, where $A^\mu=A^\mu_c t^c$ is the gluon potential in the adjoint
representation, with $t^c_{ab}=-if_{abc}$.  A transverse gauge link at the spatial infinity
is needed to retain the gauge invariance in a singular gauge~\cite{Ji:2002aa}.
\begin{figure}[t]
\begin{center}
\includegraphics[width=8cm]{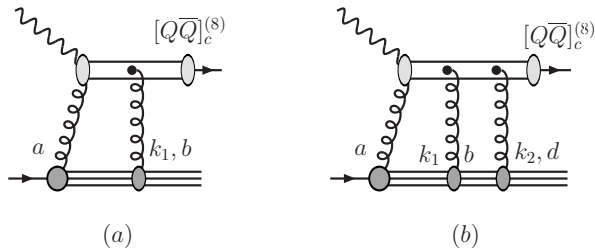}
\end{center}
\vskip -0.4cm \caption{\it Summarize the two final state
interactions in Fig.~2 into a gauge link associated with the
gluon interaction.} \label{fig3}
\end{figure}

A potential contribution to the above gluon Sivers function comes from the quark Sivers
function by splitting.
Following the calculations in~\cite{Ji:2006ub}, we find that the large $k_\perp$ gluon Sivers function can
be generated from the twist-three quark-gluon correlation function $T_F(x)$~\cite{Qiu:1991pp},
\begin{eqnarray}
&&G_{1T}^\perp=\frac{\alpha_s}{2\pi^2}\frac{\epsilon^{\alpha\beta}S_\perp^\alpha k_\perp^\beta}{(\vec{k}_\perp^2)^2}
\frac{N_c}{2}\int \frac{dx'}{x'} \left\{\left(x'\frac{\partial}{\partial x'}T_F(x',x')\right)\right.\nonumber\\
&&\times\left.(1-\xi)\left[1+(1-\xi)^2\right]+
T_F(x',x')2(1-\xi)^2\right\} \ ,
\end{eqnarray}
where $\xi=x/x'$ and a sum over all quark flavor is implicit. The overall sign of this distribution depends on the relative contribution
from the derivative and non-derivative terms, and the up and down
quark contributions.

In summary, the SSA in $ep$ collisions vanishes in the color-single model, but
survives in the color-octet model.

{\bf 3. SSA in $pp$ collisions.} Now, we turn to the SSA in heavy quarkonium
production in $pp$ collisions. In this process, we have both initial and
final state interactions. However, as we showed above, when the heavy
quark pair are produced in a color-singlet configuration, there is no final state interaction,
and we only have initial state interaction contribution. We show a typical
diagram in Fig.~4(a) in the gluon-gluon fusion subprocess. This diagram is in particular the
dominant channel for $\chi_c$ production in the color-singlet model~\cite{Bodwin:1994jh}.
The initial state interactions
for this diagram can be analyzed, and we find the SSA will depend on the gluon Sivers
function defined in Eq.~(\ref{e7}), however with a gauge link going to $-\infty$,
\begin{eqnarray}
  {\cal L}_{\xi^-,{\xi_\perp}}\to {\cal L}_\xi^{-\infty} =
  P\exp\left(-ig\int^{-\infty}_{0}
    d\zeta^- A^+(\zeta^- + \xi) \right) \ ,
\end{eqnarray}
and the associated SSA will be opposite.
\begin{figure}[t]
\begin{center}
\includegraphics[width=13cm]{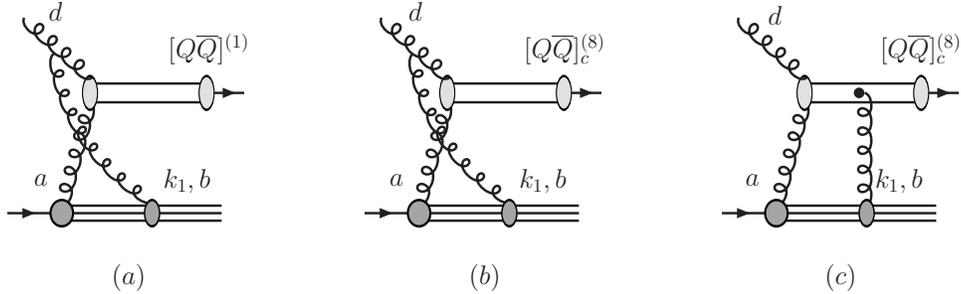}
\end{center}
\vskip -0.4cm \caption{\it Only initial state interactions
contribute to the SSA in hadron-production process in the
color-singlet model (a). On the other hand, the SSA vanishes in
the color-octet model, because the cancelation between initial
(b) and final (c) state interactions.} \label{fig4}
\end{figure}

However, if the quark pair are produced in a color-octet configuration, we will have
both initial and final state interactions. After taking into account both contributions,
the associated gauge link becomes,
$
  {\cal L}_{\xi^-,{\xi_\perp}}\to {\cal L}_\xi^{-\infty}{\cal L}_\xi^\infty
$, which is responsible for the SSA. However, the contributions from
the two gauge links cancel out each other completely.
This is because the combined gauge link is invariant under time-reversal transformation:
the two gauge links transform into each other and the combined one remains the same.
We have also checked this cancelation by an explicit calculation up to two gluon exchange contributions.
This result indicates that a standard TMD factorization breaks down for this case,
similar to recent discussions on the dijet-correlation in hadronic collisions~\cite{dijet-c}.

{\bf 4. Summary.} In this paper, we have formulated the single spin asymmetry in
heavy quarkonium production in high energy scattering in the non-relativistic limit
which should be valid in the limit $M_Q\to \infty$. Very interesting observations
were found. The SSA vanishes in $ep$ collision when the pair are
produced in a color-singlet configuration, and a nonzero SSA arises from the color-octet
contribution from the gluon Sivers function with the gauge link pointing to $+\infty$.
On the other hand, the SSA in $pp$ collisions in the color-single model depends on the
initial state interactions leading to a gluon Sivers function with gauge link pointing to
$-\infty$, and the initial and final state interactions cancel out each other in the color-octet
model and the SSA vanishes to all orders.

In our discussions, we only considered the gluon-fusion contributions. The quark channel is
also important for heavy quarkonium production, especially in $pp$ collisions at relative
low energies. In this case, the SSA will depend on the quark Sivers function from the initial state
interaction in the color-single model, which however is opposite to the Drell-Yan SSA because of
different color-factor, with a relative factor $(-1/2N_c)/C_F$. In the color-octet model, both initial
and final state interactions contribute. At one-gluon exchange order, the final state interaction
dominate over the initial one, and their total contribution  carries an overall
factor $-(N_c/2+1/2N_c)/C_F$ relative to that in the Drell-Yan process for each quark flavor
contribution. It will be interesting to compare the SSA in these two processes, and scan the
energy dependence~\cite{liu,jparc}.

We thank Xiangdong Ji, Min Liu, Jen-Chiel Peng, Ernst
Sichtermann, and Werner Vogelsang for useful discussions and
encouragements. We especially thank Jian-Ping Ma and Jianwei Qiu
for their valuable comments and discussions. This work was supported in
part by the U.S. Department of Energy under contract
DE-AC02-05CH11231. We are grateful to RIKEN, Brookhaven National
Laboratory and the U.S. Department of Energy (contract number
DE-AC02-98CH10886) for providing the facilities essential for the
completion of this work.

\end{document}